\documentclass[12pt]{article}
\usepackage{epsfig}
      \def\bp{{\bf p}}
      \def\br{{\bf r}}

      \def\D{{\cal D}}

      \def\H{{\cal H}}

      \def\R{{\cal R}}

      \def\Z{{\cal Z}}
\begin{document}

\vskip 5mm
\begin{center}
{\large\bf NUCLEAR SCISSORS WITH PAIRING AND CONTINUITY EQUATION }\\
{\large E.B. Balbutsev, L.A. Malov}\\
{\it Joint Institute for Nuclear Research, 141980 Dubna, Moscow Region,
Russia}\\
{\large P. Schuck, M. Urban}\\
\vspace*{0.2cm}
{\it Institut de Physique Nucl{\'e}aire, CNRS-IN2P3 and Universit{\'e}
Paris-Sud, 91406 Orsay Cedex, France}\\
\end{center}

\begin{abstract}
The coupled dynamics of the isovector and isoscalar giant quadrupole
resonances and low lying modes (including scissors) are studied with
the help of the Wigner Function Moments (WFM) method generalized to
take into account pair correlations.  Equations of motion for
collective variables are derived on the basis of the Time Dependent
Hartree-Fock-Bogoliubov (TDHFB) equations in the harmonic oscillator
model with quadrupole-quadrupole (QQ) residual interaction and a
Gaussian pairing force. Special care is taken of the continuity
equation.

\end{abstract}

\section{Introduction}
An exhaustive analysis of the dynamics of the scissors mode and the
isovector giant quadrupole resonance in a harmonic oscillator model
with QQ residual interaction has been performed in \cite{BaSc2}. The
WFM method was applied to derive the dynamical equations for angular
momentum and quadrupole moment. Analytical expressions for energies,
$B(M1)$- and $B(E2)$-values, sum rules and flow patterns were found
for arbitrary values of the deformation parameter. These calculations
were performed without pair correlations. However, it is well known
\cite{Zaw} that pairing is very important for the correct description
of the scissors mode. A first attempt to include pairing into the WFM
method was made in \cite{Malov}, where the description of qualitative
and quantitative characteristics of the scissors mode was drastically
improved. However, the variation of the gap during vibrations was
neglected there, resulting in a violation of the continuity equation
and in the appearance of an instability in the isoscalar channel. In
the present work we suggest a generalization of the WFM method which
takes into account pair correlations conserving the continuity
equation.

\section{Phase space moments of TDHFB equations}

The time dependent HFB equations in matrix formulation are
\cite{Solov,Ring}
\begin{equation}
i\hbar\dot\R=[\H,\R]
\label{tHFB}
\end{equation}
with
\begin{equation}
\R={\hat\rho\qquad-\hat\kappa\choose-\hat\kappa^{\dagger}\;\;1-\hat\rho^*},
\quad\H={\hat
h\quad\;\;\hat\Delta\choose\hat\Delta^{\dagger}\quad-\hat h^*}
\end{equation}
The normal density matrix $\hat \rho$ and Hamiltonian $\hat h$ are
hermitian whereas the abnormal density $\hat \kappa$ and the pairing
gap $\hat \Delta$ are skew symmetric: $\hat \kappa^{\dagger}=-\hat
\kappa^*$, $\hat \Delta^{\dagger}=-\hat \Delta^*$.

The detailed form of the TDHFB equations is
\begin{eqnarray}
i\hbar\dot{\hat\rho} =\hat h\hat\rho -\hat\rho\hat h
-\hat\Delta \hat\kappa ^{\dagger}+\hat\kappa \hat\Delta^\dagger,&&
-i\hbar\dot{\hat\kappa} =-\hat h\hat\kappa -\hat\kappa \hat h^*+\hat\Delta
-\hat\Delta \hat\rho ^*-\hat\rho \hat\Delta ,
\nonumber\\
-i\hbar\dot{\hat\rho}^*=\hat h^*\hat\rho ^*-\hat\rho ^*\hat h^*
-\hat\Delta^\dagger\hat\kappa +\hat\kappa^\dagger\hat\Delta ,&&
-i\hbar\dot{\hat\kappa}^\dagger=\hat h^*\hat\kappa^\dagger
+\hat\kappa^\dagger\hat h-\hat\Delta^\dagger
+\hat\Delta^\dagger\hat\rho +\hat\rho^*\hat\Delta^\dagger .
\label{HFB}
\end{eqnarray}
We will work with the Wigner transformation \cite{Ring} of these
equations. The relevant mathematical details can be found in
\cite{Malov}. From now on, we will not specify the spin and isospin
indices in order to make the formulae more transparent. The isospin
indices will be re-introduced at the end. In addition, we will not
write out the coordinate dependence $(\br,\bp)$ of all functions. The
Wigner transform of (\ref{HFB}) can be written as
\begin{eqnarray}
i\hbar\dot f &=&i\hbar\{h,f\}
-\Delta\kappa^{*}+\kappa\Delta^{*}
-\frac{i\hbar}{2}\{ \Delta,\kappa^{*}\}
+\frac{i\hbar}{2}\{ \kappa,\Delta^{*}\}
\nonumber\\
&&-
\frac{\hbar^2}{8}
[\{\{ \kappa,\Delta^{*}\}\}-\{\{ \Delta,\kappa^{*}\}\}]+...,
\nonumber\\
-i\hbar\dot{\bar f}&=&i\hbar\{\bar h,\bar f\}
-\Delta^{*}\kappa+\kappa^{*}\Delta
-\frac{i\hbar}{2}\{ \Delta^{*},\kappa\}
+\frac{i\hbar}{2}\{ \kappa^{*},\Delta\}
\nonumber\\
&&
+\frac{\hbar^2}{8}
[\{\{ \kappa,\Delta^{*}\}\}-\{\{ \Delta,\kappa^{*}\}\}]+...,
\nonumber\\
-i\hbar\dot\kappa&=&
-h\kappa-\kappa\bar h
-\frac{i\hbar}{2}\{h,\kappa\}
-\frac{i\hbar}{2}\{\kappa,\bar h\}
\nonumber\\
&&
+\Delta
-\Delta\bar f-f\Delta
-\frac{i\hbar}{2}\{f,\Delta\}
-\frac{i\hbar}{2}\{\Delta,\bar f\}
\nonumber\\
&&+\frac{\hbar^2}{8}
[\{\{h,\kappa\}\}+\{\{\kappa,\bar h\}\}
+\{\{ \Delta,\bar f\}\}+\{\{f,\Delta\}\}]+...,
\nonumber\\
-i\hbar\dot\kappa^{*}&=&
\kappa^{*}h+\bar h\kappa^{*}
+\frac{i\hbar}{2}\{\kappa^{*},h\}
+\frac{i\hbar}{2}\{\bar h ,\kappa^{*}\}
\nonumber\\
&&
-\Delta^{*}+\bar f\Delta^{*}+\Delta^{*} f
+\frac{i\hbar}{2}\{\bar f,\Delta^{*}\}
+\frac{i\hbar}{2}\{\Delta^{*},f\}
\nonumber\\
&&
-\frac{\hbar^2}{8}
[\{\{\kappa^{*},h\}\}+\{\{\bar h ,\kappa^{*}\}\}
+\{\{\bar f,\Delta^{*}\}\}+\{\{\Delta^{*},f\}\}]+... ,
\label{WHFB}
\end{eqnarray}
where the functions $h$, $f$, $\Delta$, and $\kappa$ are the Wigner
transforms of $\hat h$, $\hat\rho$, $\hat\Delta$, and $\hat\kappa$,
respectively, $\bar f(\br,\bp)=f(\br,-\bp)$, $\{f,g\}$ is the Poisson
bracket of the functions $f(\br,\bp)$ and $g(\br,\bp)$ and
$\{\{f,g\}\}$ is their double Poisson bracket;
the dots stand for terms proportional to higher powers of $\hbar$.

In order to study collective modes described by these equations, we
apply the method of Wigner function moments (or phase space
moments). The idea of the method is based on the virial theorems by
Chandrasekhar and Lebovitz \cite{Chand}; its detailed formulation can
be found in \cite{Bal,BaSc}. For the investigation of the quadrupole
collective motion with $K^{\pi}=1^+$ in axially symmetric nuclei, it is
necessary to calculate moments of Eqs.~(\ref{WHFB}) with the weight
functions
\begin{equation}
W = \quad xz,\quad p_xp_z,\quad zp_x+xp_z\equiv \hat L,\quad
\mbox{and}\quad zp_x-xp_z\equiv\hat I_y.
\label{weightfunctions}
\end{equation}
This procedure means that we refrain from seeking the whole density
matrix and restrict ourselves to the knowledge of only several
moments. Nevertheless, this information turns out to be sufficient for
a satisfactory description of various collective modes with quantum
numbers $K^{\pi}=1^+$, as it was shown in our previous publications
\cite{BaSc2,Bal,BaSc}. In the case without pairing, this restricted
information can be extracted from the TDHF equations and becomes exact
only for the harmonic oscillator with multipole-multipole residual
interactions. For more realistic models it becomes approximate even
without pairing. The TDHFB equations (\ref{WHFB}) are considerably
more complicated than the TDHF ones, so additional approximations are
necessary even for the simple model considered here, as will be
discussed below.

Integration of Eqs.~(\ref{WHFB}) (including the terms of higher orders
in $\hbar$) over the phase space with the weight $W$, where $W$ is any
one of the weight functions listed in (\ref{weightfunctions}), yields
the following set of equations:
\begin{eqnarray}
      i\hbar\frac{d}{dt}\int\! d(\bp,\br) W f=
\int\! d(\bp,\br)
\left[i\hbar \{W,h\}f+W(\Delta^{*}\kappa-\kappa^{*}\Delta)
-\frac{i\hbar}{2}(\{W,\Delta^{*}\}\kappa+\{W,\Delta\}\kappa^{*})
\right.
\nonumber\\
\left.
-\frac{\hbar^2}{8}
(\{\{W,\Delta^{*}\}\}\kappa-\{\{W,\Delta\}\}\kappa^{*})
\right],
\nonumber\\
     i\hbar\frac{d}{dt}\int\! d(\bp,\br) W\bar f=
\int\! d(\bp,\br)
\left[i\hbar\{\bar h,W\}\bar f+W(\Delta^{*}\kappa-\kappa^{*}\Delta)
+\frac{i\hbar}{2}(\{W,\Delta^{*}\}\kappa+\{W,\Delta\}\kappa^{*})
\right.
\nonumber\\
\left.
-\frac{\hbar^2}{8}
(\{\{W,\Delta^{*}\}\}\kappa-\{\{W,\Delta\}\}\kappa^{*})
\right],
\nonumber\\
     i\hbar\frac{d}{dt}\int\! d(\bp,\br) W\kappa=
\int\! d(\bp,\br)\left[W(h+\bar h)\kappa
+\frac{i\hbar}{2}\{W,(h-\bar h)\}\kappa
-W\Delta(1-\bar f-f)
\right.
\hspace{16mm}
\nonumber\\
\left.
+\frac{i\hbar}{2}\{W,\Delta\}(\bar f-f)
-\frac{\hbar^2}{8}
[\{\{W,(h+\bar h)\}\}\kappa
+\{\{W,\Delta\}\}(\bar f+f)]
\right],
\nonumber\\
     i\hbar\frac{d}{dt}\int\! d(\bp,\br) W\kappa^{*}=
\int\! d(\bp,\br)\left[-W(h+\bar h)\kappa^{*}
+\frac{i\hbar}{2}\{W,(h-\bar h)\}\kappa^{*}
+W\Delta^{*}(1-\bar f-f)
\right.
\nonumber\\
\left.
+\frac{i\hbar}{2}\{W,\Delta^{*}\}(\bar f-f)
+\frac{\hbar^2}{8}
[\{\{W,(h+\bar h)\}\}\kappa^{*}
+\{\{W,\Delta^{*}\}\}(\bar f+f)]
\right],
\label{HFBW}
\end{eqnarray}
where $\int\! d(\bp,\br)\equiv
2(2\pi\hbar)^{-3}\int\!d^3p\,\int\!d^3r$. It is necessary to note an
essential point: there are no terms with higher powers of $\hbar$ in
these equations. The infinite number of terms proportional to
$\hbar^n$ with $n>2$ have disappeared after integration. This fact
does not mean that higher powers of $\hbar$ are not necessary for the
exact solution of the problem. As it will be shown below, the
equations (\ref{HFBW}) contain terms which are coupled to dynamical
equations for higher-order moments, which include, naturally,
higher powers of $\hbar$.

\subsection{Continuity equation}

The moments of the first two equations with the weight $xz$ are of
special interest, because in this case equations are integrated over
momentum space without any weight. It is known that the integration of
the Vlasov (or Boltzmann) equation over $\bp$ gives the continuity
equation \cite{Ring}; the same is true for the Wigner function
equation. The inclusion of pair correlations must not destroy this
property of the Wigner function, so one can expect that integration
over $\bp$ of the first (and second) equation in (\ref{WHFB}) will
produce the continuity equation. Though this is known from general
arguments, let us repeat it in detail.

First of all we integrate over $\bp$ the first part of the first
equation in (\ref{WHFB}):
\begin{equation}
\int\! d^3p \dot f=\int\! d^3p\sum_{i=1}^3\left(
\frac{\partial h}{\partial x_i}\frac{\partial f}{\partial p_i}-
\frac{\partial h}{\partial p_i}\frac{\partial f}{\partial x_i}\right)
  + \cdots \quad \mbox{(terms containing $\Delta$ and $\kappa$)}.
\label{Con1}
\end{equation}
In the case of a velocity-independent potential, the first integral on
the right-hand side is equal to zero. By definition $\int\! d^3p
f(\br,\bp)=n(\br)$ and $\int\! d^3p p_if(\br,\bp)=mn(\br)u_i(\br)$,
where $n(\br)$ is the density of particles and $u_i(\br)$ is the
$i$-th component of their mean velocity. Using $\partial h/\partial
p_i = p_i/m$, one obtains
\begin{equation}
\dot n+div(n{\bf u})= \mbox{terms containing $\Delta$ and $\kappa$}.
\end{equation}
So, we will recover the continuity equation if the integral over $\bp$
of all the terms of the first equation in (\ref{WHFB}) containing
$\Delta$ and $\kappa$ gives zero.

Let us integrate over $\bp$ the term $\kappa\Delta^* -
\kappa^*\Delta$. It is known that the pairing gap and the abnormal
density are connected by the integral relation (see, e.g.,
ref. \cite{Ring}):
\begin{equation}
\Delta(\br,\bp)=-\int\! \frac{d^3p'}{(2\pi\hbar)^3}
v(|\bp-\bp'|)\kappa(\br,\bp').
\label{DK}
\end{equation}
Using this relation, one finds
\begin{equation}\int\!d^3p(\kappa\Delta^*-\kappa^*\Delta)=
\int\!d^3p\int\!d^3p'[\kappa(\br,\bp)\kappa^*(\br,\bp')
-\kappa^*(\br,\bp)\kappa(\br,\bp')]v(|\bp-\bp'|)=0.
\end{equation}
The equality becomes obvious after changing the variables
$\bp\leftrightarrow \bp'$ in the second (or first) part of this
formula.

Finally, let us integrate over $\bp$ the term with Poisson brackets:
\begin{eqnarray}
\int\!d^3p\int\!d^3p'\sum_{i=1}^3\left[
\frac{\partial\kappa^*(\br,\bp)}{\partial x_i}
\frac{\partial v(|\bp-\bp'|)}{\partial p_i}\kappa(\br,\bp')
-\frac{\partial\kappa^*(\br,\bp)}{\partial p_i}v(|\bp-\bp'|)
\frac{\partial\kappa(\br,\bp')}{\partial x_i}\right.
\nonumber\\
+\left.\frac{\partial\kappa(\br,\bp)}{\partial x_i}
\frac{\partial v(|\bp-\bp'|)}{\partial p_i}\kappa^*(\br,\bp')
-\frac{\partial\kappa(\br,\bp)}{\partial p_i}v(|\bp-\bp'|)
\frac{\partial\kappa^*(\br,\bp')}{\partial x_i}\right]
\nonumber\\
=\int\!d^3p\int\!d^3p'\sum_{i=1}^3
\left[\frac{\partial\kappa^*(\br,\bp)}{\partial x_i}
\kappa(\br,\bp')
+\kappa^*(\br,\bp)
\frac{\partial\kappa(\br,\bp')}{\partial x_i}\right.
\nonumber\\
+\left.\frac{\partial\kappa(\br,\bp)}{\partial x_i}
\kappa^*(\br,\bp')
+\kappa(\br,\bp)
\frac{\partial\kappa^*(\br,\bp')}{\partial x_i}\right]
\frac{\partial v(|\bp-\bp'|)}{\partial p_i}=0.
\label{Pois}
\end{eqnarray}
The last equality becomes obvious after changing the variables
$p\leftrightarrow p'$ and using the relation $\partial
v(|\bp-\bp'|)/\partial p_i = -\partial v(|\bp-\bp'|)/\partial p'_i$.
In a similar way one can show that integration over $\bp$ of any term
of higher order in $\hbar$ will give zero.

So, as expected, one can conclude that pairing does not spoil the
continuity equation, which is contained in the TDHFB equations. As a
result, the $\Delta$-dependent terms in the first and second equations
of (\ref{HFBW}) disappear, when the weight $W$ does not depend on
$\bp$, for example, $W=xz$ (see below).

\subsection{Linearization}

It is convenient to rewrite the equations (\ref{HFBW}) in terms of
$h_{\pm}=h\pm \bar h$, $f_{\pm}=f\pm\bar f$,
$\Delta_{\pm}=\Delta\pm\Delta^*$, $\kappa_{\pm}=\kappa\pm\kappa^*$.
These equations are strongly nonlinear.  Having in mind small
amplitude oscillations, we will linearize: $f_{\pm}=f_{\pm}^{0}+\delta
f_{\pm}$, $\kappa_{\pm}=\kappa^{0}_{\pm}+\delta\kappa_{\pm}$,
$\Delta_{\pm}=\Delta^{0}_{\pm}+\delta\Delta_{\pm}$.  The Hamiltonian
should be divided into the ground state Hamiltonian $h^{0}$ and the
residual interaction $h^{1}$ (and, if necessary, the external
field). We consider $h^{0}$ without $\bp$-odd terms, hence $h^{0}_-=0$
and as a consequence $f_-^{0}=0$. It is natural to take $\Delta^{0}$
real, i.e.  $\Delta^{0}_-=0$, $\kappa^{0}_-=0$ and
$\Delta_+^0=2\Delta^0$, $\kappa_+^0=2\kappa^0$.  Linearizing
(\ref{HFBW}), we arrive at
\begin{eqnarray}
&&      i\hbar\frac{d}{dt}\int\! d(\bp,\br) W \delta f_+=
\int\! d(\bp,\br)\left[
\frac{i\hbar}{2}(
\{W,h_{+}^{0}\}\delta f_-
+\{W,h_{-}^{1}\}f_+^{0})
+W(\Delta_+^{0}\delta \kappa_--\kappa_+^{0}\delta \Delta_-)
\right.\hspace*{5mm}
\nonumber\\
&&\hspace*{40mm}
\left.
-\frac{\hbar^2}{8}
(\{\{W,\Delta_+^{0}\}\}\delta \kappa_-
-\{\{W,\kappa_+^{0}\}\}\delta \Delta_-)
\right],
\nonumber\\
&&      i\hbar\frac{d}{dt}\int\! d(\bp,\br) W \delta f_-=
\int\! d(\bp,\br)\left[
\frac{i\hbar}{2}(
\{W,h_{+}^{0}\}\delta f_+
+\{W,h_{+}^{1}\}f_+^{0})
\right.
\nonumber\\
&&\hspace*{40mm}
\left.
-\frac{i\hbar}{2}(\{W,\Delta_+^{0}\}\delta \kappa_+
-\{W,\kappa_+^{0}\}\delta \Delta_+)
\right],
\nonumber\\
&&i\hbar\frac{d}{dt}\int\! d(\bp,\br) W\delta \kappa_+=
\int\! d(\bp,\br)\left[Wh_+^{0}\delta \kappa_-
+\frac{i\hbar}{2}\{W,h_-^{1}\}\kappa_+^{0}
-W\delta \Delta_-(1-f_+^{0})
\right.
\nonumber\\
&&\hspace*{40mm}
\left.
-\frac{i\hbar}{2}\{W,\Delta_+^{0}\}\delta f_-
-\frac{\hbar^2}{8}
(\{\{W,h_+^{0}\}\}\delta \kappa_-
+\{\{W,f_+^0\}\}\delta \Delta_-)
\right],
\nonumber\\
&&i\hbar\frac{d}{dt}\int\! d(\bp,\br) W\delta \kappa_-=
\int\! d(\bp,\br)\left[Wh_+^{0}\delta \kappa_+
+Wh_+^{1}\kappa_+^{0}
-W\delta \Delta_+(1-f_+^{0})
+W\Delta_+^{0}\delta f_+
\right.
\nonumber\\
&&\hspace*{20mm}
\left.
-\frac{\hbar^2}{8}(\{\{W,h_+^{0}\}\}\delta \kappa_+
+\{\{W,h_+^{1}\}\}\kappa_+^{0}
+\{\{W,\Delta_+^{0}\}\}\delta f_+
+\{\{W,f_+^{0}\}\}\delta \Delta_+)
\right].
\label{HFBWlin}
\end{eqnarray}

Let us look more closely at the variation of the gap,
$\delta\Delta(\br,\bp)$, which should be expressed in terms of
$\delta\kappa(\br,\bp)$. In the framework of the method of moments,
this can be done quite easily. According to (\ref{DK}), the relevant
variations are connected by the integral relation
\begin{equation}
\delta\Delta(\br,\bp)=-\int\! \frac{d^3p'}{(2\pi\hbar)^3}
v(|\bp-\bp'|)\delta\kappa(\br,\bp').
\label{varDK}
\end{equation}
Substituting this into (\ref{HFBWlin}) and changing the variables
$\bp\leftrightarrow \bp'$, we obtain immediately the desired result.
Examples for this type of calculation can be found in Appendix A.

Until this point, our formulation is completely general.  To proceed
further, we are forced to make some approximations to get rid of
higher-rank moments and to obtain a closed set of dynamical equations
for the second-rank moments. The usual problems of the method of
moments are connected with integrals of the type $\int\!
d(\bp,\br)W\Delta^0\delta\kappa$, $\int\!
d(\bp,\br)\{W,\Delta^0\}\delta\kappa$, $\int\!
d(\bp,\br)Wf^0\delta\Delta$, $\int\! d(\bp,\br)\{W,\Delta^0\}\delta f$
etc.  The variations $\delta\kappa(\br,\bp)$, $\delta f(\br,\bp)$ are
integrated here with weights which are more complicated functions of
$\br, \bp$ than the simple functions $W$. So the problem arises: how
to express these integrals via the moments $\int\!
d(\bp,\br)W\delta\kappa(\br,\bp)$, $\int\! d(\bp,\br)W\delta
f(\br,\bp)$ which we work with? To solve this problem, one can develop
the functions $f^0(\br,\bp)$, $\kappa^0(\br,\bp)$, $\Delta^0(\br,\bp)$
in a Taylor series around some point $(\bar \br,\bar \bp)$. However,
another problem appears on this way: what to do with the higher-order
moments which will inevitably be generated by the Taylor series? We
suggest to neglect them, because it is natural to expect that the
influence of higher-order moments on the dynamics of lower-order
moments will be small \cite{Bal}.

A few words about the choice of the point $(\bar \br,\bar \bp)$. It is
known that all the dynamics happens in the vicinity of the Fermi
surface. Therefore, the choice of the momentum $\bar\bp$ is obvious:
it should be equal to the Fermi momentum $\bp_F(\bar \br)$. The choice
of $\bar\br$ is more complicated, because it depends on the nature of
the mode under consideration. For example, in the case of surface
vibrations, $\bar\br$ should be taken somewhere near the nuclear
surface ${\bf R}$. In the case of compressional modes, it is more
appropriate to take $\bar\br$ somewhere inside the nucleus. In any
case, it is a rather delicate problem and every particular example
requires the careful analysis. In principle, the value of $\bar\br$
can be used as a fitting parameter.

\section{Equations of motion}

The model considered here is a harmonic oscillator with
quadrupole-quadrupole residual interaction \cite{BaSc2}. The
corresponding mean-field Hamiltonians for protons and neutrons are
\begin{equation}
 h^{\tau}=\frac{p^2}{2m}+\frac{1}{2}m
\omega^2r^2+\sum_{\mu=-2}^2(-1)^{\mu}B_{2\mu}^{\tau}(t)q_{2\mu}(\br)
-\mu^{\tau},
\label{htau}
\end{equation}
where $B_{2\mu}^{\rm n}=\kappa_{nn}Q_{2\mu}^{\rm n}
+\kappa_{np}Q_{2\mu}^{\rm p}$, $B_{2\mu}^{\rm
p}=\kappa_{pp}Q_{2\mu}^{\rm p} +\kappa_{np}Q_{2\mu}^{\rm n}$,
$q_{2\mu}(\br)=\sqrt{16\pi/5}r^2Y_{2\mu}(\theta,\phi)$, $\mu^{\tau}$
being the chemical potential of protons ($\tau$=p) or neutrons
($\tau$=n). $\kappa_{\tau\tau'}$ is the strength constant, and we
suppose $\kappa_{\rm nn}=\kappa_{\rm pp}$. $Q_{2\mu}^{\tau}$ is a component of
the quadrupole moment
\begin{equation}
Q_{2\mu}^{\tau}(t)
=\int\! d(\bp,\br) q_{2\mu}(\br) f^{\tau}(\br,\bp,t)
=\int\! d(\bp,\br) q_{2\mu}(\br) [f_0^{\tau}(\br,\bp)
+\delta f^{\tau}(\br,\bp,t)]=Q_{2\mu}^{\tau0}+\delta Q_{2\mu}^{\tau}(t).
\end{equation}
The Hamiltonian $h^{\tau}$ is divided into the equilibrium part
$h_0^{\tau}$ and the variation $h_1^{\tau}$:
\begin{eqnarray}
&& h_0^{\tau}=\frac{p^2}{2m}+\frac{1}{2}m
\omega^2r^2+B_{20}^{\tau 0}(t)q_{20}(\br)-\mu^{\tau}
=\frac{p^2}{2m}+\frac{1}{2}m
[\omega_x^{\tau2}(x^2+y^2)+\omega_z^{\tau2}z^2]-\mu^{\tau},
\nonumber\\
&&h_1^{\tau}=\sum_{\mu=-2}^2(-1)^{\mu}\delta B_{2\mu}^{\tau}(t)q_{2\mu}(\br),
\label{h01}
\end{eqnarray}
where we took into account that in an axially symmetric nucleus
$Q_{2\pm1}^{\tau0}=Q_{2\pm2}^{\tau0}=0$.
Obviously, in this
model $h_-^{\tau}=0$ and $h_+^{\tau}=2h^{\tau}$.

The required Poisson brackets together with the definition of
$\omega_x^{\tau}$ and $\omega_z^{\tau}$ are written out in appendix B.
For the sake of simplicity, we will neglect in (\ref{HFBWlin}) all
terms proportional to $\hbar^2$ (quantum corrections) except the
simplest one, namely the term with $h^1_+$, which we will keep in
order to have an idea about the possible influence of quantum
corrections. We assume that $\Delta^0$ does not depend on the
direction of $\bp$, i.e. $\Delta^0(\br,\bp)=\Delta(\br,p)$ (see
Appendix A). Let us introduce the collective variables
\begin{eqnarray}
Q^{\tau}_{\pm}(t)=\int\! d(\bp,\br) xz \delta
f^{\tau}_{\pm}(\br,\bp,t),\quad&&
\tilde Q^{\tau}_{\pm}(t)=\int\! d(\bp,\br) xz \delta
\kappa^{\tau}_{\pm}(\br,\bp,t),
\nonumber\\
P^{\tau}_{\pm}(t)=\int\! d(\bp,\br) p_xp_z \delta
f^{\tau}_{\pm}(\br,\bp,t),\quad&&
\tilde P^{\tau}_{\pm}(t)=\int\! d(\bp,\br) p_xp_z \delta
\kappa^{\tau}_{\pm}(\br,\bp,t),
\nonumber\\
L^{\tau}_{\pm}(t)=\int\! d(\bp,\br)\hat L \delta
f^{\tau}_{\pm}(\br,\bp,t),\quad&&
\tilde L^{\tau}_{\pm}(t)=\int\! d(\bp,\br)\hat L \delta
\kappa^{\tau}_{\pm}(\br,\bp,t),
\nonumber\\
I_{\pm}^{\tau}(t)=\int\! d(\bp,\br)\hat I_y \delta
f^{\tau}_{\pm}(\br,\bp,t),\quad&&
\tilde I^{\tau}_{\pm}(t)=\int\! d(\bp,\br)\hat I_y \delta
\kappa^{\tau}_{\pm}(\br,\bp,t).
\label{Varis}
\end{eqnarray}
By definition, $L_+=I_+=P_-=Q_-=0.$ The analysis of the dynamical
equations shows that $\tilde L_+=\tilde I_+=\tilde L_-=\tilde I_-=0,$
and we are left with the following eight equations:
\begin{eqnarray}
1.\hspace{5mm}i\hbar \dot Q_+&=&i\hbar \frac{1}{m}L_-,
\nonumber\\
2.\hspace{5mm}i\hbar \dot{\tilde Q}_+&=&(2h_0-|V_0|I^{f\Delta}_{xz})
\tilde Q_-
+\frac{i\hbar}{2} \frac{|V_0|}{\hbar^2}I_{\partial p}L_-,
\nonumber\\
3.\hspace{5mm}i\hbar \dot{\tilde Q}_-&=&(2h_0-|V_0|I^{f\Delta}_{xz})
\tilde Q_+
+2k_4\Z(t)+2\Delta^0 Q_+,
\nonumber\\
4.\hspace{5mm}i\hbar \dot P_+&=&-i\hbar \frac{m}{2}
[\omega_+^2L_--\omega_-^2I_-]+|V_0|I^{\kappa\Delta}_{pp}\tilde P_-,
\nonumber\\
5.\hspace{5mm}i\hbar \dot{\tilde P}_+&=&(2h_0-|V_0|I^{f\Delta}_{pp})
\tilde P_-
-\frac{i\hbar}{2} |V_0|I_{\partial r}[\omega_+^2L_-
-\omega_-^2I_-],
\nonumber\\
6.\hspace{5mm}i\hbar \dot{\tilde P}_-&=&(2h_0-|V_0|I^{f\Delta}_{pp})
\tilde P_+
+2\Delta^0 P_+ -\frac{\hbar^2}{2}k_0\Z(t),
\nonumber\\
7.\hspace{5mm}i\hbar \dot L_-&=&i\hbar \left[\frac{2}{m}P_+
-m\omega_+^2Q_+-2(\langle z^2\rangle+\langle x^2\rangle)\Z(t)
+\frac{|V_0|}{\hbar^2}I_{xp}^{\{\kappa\Delta\}}\tilde P_+\right],
\nonumber\\
8.\hspace{5mm}i\hbar \dot I_-&=&-i\hbar [m\omega_-^2Q_+
+2(\langle z^2\rangle-\langle x^2\rangle)\Z(t)].
\label{HFB8}
\end{eqnarray}
Here $\Z^{\tau}(t)=12\beta_{13}^{\tau}(t)$ (see Appendix B) with
$\beta_{13}^{\rm n}=\frac{1}{2}[\kappa Q_{+}^{\rm n}+\bar\kappa
Q_{+}^{\rm p}]$ and $\beta_{13}^{\rm p}=\frac{1}{2}[\kappa Q_{+}^{\rm
p}+\bar\kappa Q_{+}^{\rm n}]$, $\langle x^2\rangle= \int\!
d(\bp,\br)x^2f_0(\br,\bp)$, $\langle z^2\rangle= \int\!
d(\bp,\br)z^2f_0(\br,\bp)$, $k_0=2\int\! d(\bp,\br)
\kappa^{0}(\br,\bp)$, $k_4=2\int\! d(\bp,\br)x^2z^2
\kappa^{0}(\br,\bp)$ and $\omega_{\pm}^2=\omega_x^2 \pm \omega_z^2$.
The integrals $I^{f\Delta}_{xz}$, $I^{f\Delta}_{pp}$,
$I^{\kappa\Delta}_{pp}$, $I^{\{\kappa\Delta\}}_{xp}$, $I_{\partial p}$
and $I_{\partial r}$ are defined in Appendix A. All coefficients
should be calculated at the point $(\bar\br,\bp_F(\bar\br))$, for
example, $\Delta^0=\Delta^0(\bar\br,\bp_F(\bar\br))$. Note that the
first equation does not contain $\Delta$-dependent terms. It has the
typical structure \cite{BaSc2,Bal} which is characteristic for
coordinate moments of the continuity equation. The last equation does
not depend on pairing parameters, either. However, this is due to the
symmetry property of the operator $\hat I$ (see (\ref{IL})) and has
nothing to do with the continuity equation.

Let us compare these equations with the respective equations of
\cite{Malov}, which were derived using the approximation
\begin{equation}
\Delta^0=const,\quad \delta\Delta(\br,\bp,t)=0,
\label{approx0}
\end{equation}
resulting in a violation of the continuity equation. Written in terms
of the variables (\ref{Varis}), they read
\begin{eqnarray}
   i\hbar \dot Q_+&=&i\hbar \frac{1}{m}L_- +2\Delta^0\tilde Q_-,
\nonumber\\
   i\hbar \dot{\tilde Q}_-&=&
2\Delta^0 Q_+ +2k_4\Z(t),
\nonumber\\
   i\hbar \dot P_+&=&-i\hbar \frac{m}{2}
[\omega_+^2L_- -\omega_-^2I_-]
+2\Delta^0\tilde P_-,
\nonumber\\
   i\hbar \dot{\tilde P}_-&=&
2\Delta^0 P_+ -\frac{\hbar^2}{2}k_0\Z(t),
\nonumber\\
   i\hbar \dot L_-&=&i\hbar [\frac{2}{m}P_+
 -m\omega_+^2Q_+
-2(\langle z^2\rangle+\langle x^2\rangle)\Z(t)],
\nonumber\\
   i\hbar \dot I_-&=&-i\hbar [m\omega_-^2Q_+
+2(\langle z^2\rangle-\langle x^2\rangle)\Z(t)].
\label{HFB0}
\end{eqnarray}
The most evident difference between the sets of equations (\ref{HFB8})
and (\ref{HFB0}) is the number of equations: eight and six,
respectively. How could this happen? The approximation
(\ref{approx0}), used in \cite{Malov}, makes the third equation of
(\ref{HFBWlin}) trivial (i.e. its right-hand side becomes equal to
zero identically), producing two integrals of motion $\dot{\tilde
Q_+}=0$ and $\dot{\tilde P_+}$, which should be included in the set
(\ref{HFB0}).

The second and the most important difference between the two systems
of equations concerns the first equations of (\ref{HFB8}) and
(\ref{HFB0}). The dynamical equation for the variable $Q_+$ in
(\ref{HFB0}) contains the additional (in comparison with (\ref{HFB8}))
term $2\Delta^0\tilde Q_-$, whose existence is a direct consequence of
the violation of the continuity equation. This is the only place where
the violation of the continuity equation appears explicitly. There are
more differences between (\ref{HFB8}) and (\ref{HFB0}), which are all
connected with the approximation (\ref{approx0}). For example, the
dynamical equation for $P_+$ in (\ref{HFB8}) has the term
$|V_0|I_{pp}^{\kappa\Delta}\tilde P_-$ instead of the corresponding
term $2\Delta^0\tilde P_-$ in (\ref{HFB0}). The integral
$I_{pp}^{\kappa\Delta}$ contains the contributions from $\delta\kappa$
(the factor $2\Delta^0$) and from $\delta\Delta$ as well (see Appendix
A, formulae (\ref{Int1}) and (\ref{I11})).

And the last difference: the set of equations (\ref{HFB0}) has the
pleasant property that its eigenmodes can be found analytically,
contrary to those of the set (\ref{HFB8}). There is no necessity to
explain how important and convenient it is to have (even approximate)
analytical solutions of a problem. It turns out that one can find an
approximation which allows one to get analytical solutions of
(\ref{HFB8}) without violating the continuity equation.

From general arguments one can expect that the phase of $\Delta$ (and
of $\kappa$, since both are linked according to equation (\ref{DK}))
is much more relevant than its magnitude, since the former determines
the superfluid velocity. After linearization, the phase of $\Delta$
($\kappa$) is expressed by $\delta\Delta_-$ ($\delta\kappa_-$), while
$\delta\Delta_+$ ($\delta\kappa_+$) describes oscillations of the
magnitude of $\Delta$ ($\kappa$). Let us therefore assume that
\begin{equation}
\delta\kappa_+(\br,\bp)\ll\delta\kappa_-(\br,\bp).
\label{approx1}
\end{equation}
This assumption was explicitly confirmed in ref. \cite{Urban} for the case
of superfluid trapped fermionic atoms, where it was shown that
$\delta\Delta_+$ is suppressed with respect to $\delta\Delta_-$ by one
order of $\Delta/E_F$, where $E_F$ denotes the Fermi energy.

The assumption (\ref{approx1}) does not contradict the equations of
motion and allows one to neglect all terms containing the variables
$\tilde Q_+$ and $\tilde P_+$ in the equations No. 3, 6, and 7 of
(\ref{HFB8}). In this case the "small" variables $\tilde Q_+$, $\tilde
P_+$ will not affect the dynamics of the six "big" variables $Q_+$,
$P_+$, $L_-$, $I_-$, $\tilde Q_-$, $\tilde P_-$. This means that the
dynamical equations for the "big" variables can be considered
independently from that of the "small" variables, and we will finally
deal with a set of only six equations. Adding the isospin index
$\tau=$(n,p), we have
\begin{eqnarray}
   i\hbar \dot Q_+^{\tau}&=&i\hbar \frac{1}{m}L_-^{\tau},
\nonumber\\
   i\hbar \dot{\tilde Q}_-^{\tau}&=&
2\Delta^{0\tau} Q_+^{\tau}+2k_4^{\tau}\Z^{\tau}(t),
\nonumber\\
   i\hbar \dot P_+^{\tau}&=&-i\hbar \frac{m}{2}
[\omega_+^{\tau2}L_-^{\tau}-\omega_-^{\tau2}I_-^{\tau}]
+|V_0^{\tau}|(I^{\kappa\Delta}_{pp})^{\tau}\tilde P_-^{\tau},
\nonumber\\
   i\hbar \dot{\tilde P}_-^{\tau}&=&
2\Delta^{0\tau} P_+^{\tau} -\frac{\hbar^2}{2}k_0^{\tau}\Z^{\tau}(t),
\nonumber\\
   i\hbar \dot L_-^{\tau}&=&i\hbar [\frac{2}{m}P_+^{\tau}
 -m\omega_+^{\tau2}Q_+^{\tau}
-2(\langle z^2\rangle^{\tau}+\langle x^2\rangle^{\tau})\Z^{\tau}(t)],
\nonumber\\
   i\hbar \dot I_-^{\tau}&=&-i\hbar [m\omega_-^{\tau2}Q_+^{\tau}
+2(\langle z^2\rangle^{\tau}-\langle x^2\rangle^{\tau})\Z^{\tau}(t)].
\label{HFB6}
\end{eqnarray}
It is interesting to compare the results of the two different
approximations: the sets of equations (\ref{HFB6}) and
(\ref{HFB0}). The difference is minor: the factor
$|V_0|I^{\kappa\Delta}_{pp}$ in the third equation of (\ref{HFB6})
instead of $2\Delta^0$ in (\ref{HFB0}) and the absence of the term
$2\Delta^0\tilde Q_-$ in the first equation of (\ref{HFB6}) contrary
to (\ref{HFB0}). Calculations show that numerically the factor
$|V_0|I^{\kappa\Delta}_{pp}$ is not so far from $2\Delta^0$ (see
fig. 1), so one can conclude that the approximations (\ref{approx1})
and (\ref{approx0}) lead to similar dynamical equations. On the other
hand, the approximation (\ref{approx1}) is undoubtedly better than
(\ref{approx0}), because it is physically motivated and it does not
violate the continuity equation.

In this paper we will treat the reduced set of equations (\ref{HFB6})
that can be solved analytically. This allows us to compare our results
with those of \cite{Malov} and to assess the quantitative effect of
the correct treatment of the continuity equation. The numerical
analysis of the full problem (\ref{HFB8}) with all coefficients
calculated in the microscopic approach will be postponed to a future
publication.

\section{Analytical solution}

First of all, we rewrite the equations (\ref{HFB6}) in terms of the
isovector and isoscalar variables $\underline Q_{\pm}=Q_{\pm}^{\rm
n}-Q_{\pm}^{\rm p}$, $Q_{\pm}=Q_{\pm}^{\rm n}+Q_{\pm}^{\rm
p}$, $\underline {\tilde Q}_{\pm}=\tilde Q_{\pm}^{\rm n}-
\tilde Q_{\pm}^{\rm p}$, $\tilde Q_{\pm}=\tilde Q_{\pm}^{\rm n} +\tilde
Q_{\pm}^{\rm p}$, and so on. In order to separate the isovector and
isoscalar sets of equations, we employ the standard approximation
which works very well in the case of collective motion \cite{Bal}:
\begin{equation}
Q_+^{\rm n}/N=\pm Q_+^{\rm p}/Z,
\label{isoappr}
\end{equation}
where $N$ ($Z$) is the number of neutrons (protons) and the sign +
($-$) is utilized for the isoscalar (isovector) motion.

\subsection{Isovector eigenfrequencies}

The set of equations describing isovector excitations reads
\begin{eqnarray}
i\hbar \underline {\dot Q}_+&=&i\hbar \frac{1}{m}\underline L_-,
\nonumber\\
i\hbar \underline {\dot{\tilde Q}}_-&=&
(\chi_0k_4'+2\Delta')\underline Q_+,
\nonumber\\
i\hbar \underline {\dot P}_+&=&
2\tilde\Delta'\underline {\tilde P}_-
-i\hbar m\bar\omega^2
\{[1+\frac{\delta}{3}(1-\alpha\xi)]\underline L_-
-\delta(1-\alpha\xi) \underline I_-\},
\nonumber\\
i\hbar \underline {\dot{\tilde P}}_-&=&
2\Delta'\underline P_+
-\frac{\hbar^2}{4}\chi_0k_0'\underline Q_+,
\nonumber\\
i\hbar \underline {\dot L}_-&=&i\hbar \frac{2}{m}\underline P_+
-i\hbar2m\bar\omega^2\psi
\underline Q_+,
\nonumber\\
i\hbar \underline {\dot I}_-&=&
-i\hbar\delta(1-\alpha)(1-\xi)2m\bar\omega^2\underline Q_+.
\label{Ivec}
\end{eqnarray}

The following notations are introduced here: $\chi_0=12\kappa_0,$
$\,\psi=[(1+\frac{\delta}{3})(1-\alpha-\xi)
-\frac{\delta}{3}\alpha\xi],$ $\,k_i'=\alpha k_i^++(\nu-\pi)k_i^-,$
$\,k_i^{\pm}=k_i^{\rm n}\pm k_i^{\rm p},$ $\,i=0,4,$
$\,\Delta'=\nu\Delta^{\rm n}+\pi\Delta^{\rm p},$
$\,\tilde\Delta'=\nu\tilde\Delta^{\rm n}+\pi \tilde\Delta^{\rm p},$
$2\tilde\Delta^{\tau}=|V_0|(I^{\kappa\Delta}_{pp})^{\tau}$, $\nu=N/A$,
$\pi=Z/A$, $A=N+Z$, $\xi=(\nu-\pi)(Q_{00}^{\rm n}-Q_{00}^{\rm
p})/Q_{00}$, $Q_{00}=Q_{00}^{\rm n}+Q_{00}^{\rm p}$.  We use the
standard \cite{BM} definition of the deformation parameter
$\delta=3Q_{20}/(4Q_{00})$, where $Q_{00}=\int\! d(\bp,\br)
(x^2+y^2+z^2)f^0(\br,\bp) =2\langle x^2\rangle+\langle z^2\rangle$ and
$Q_{20}=2\langle z^2\rangle-2\langle x^2\rangle$ are monopole and
quadrupole moments, respectively.  Usually one takes the isovector
strength constant $\kappa_1=\frac{1}{2}(\kappa_{nn}-\kappa_{np})$
proportional to the isoscalar one,
$\kappa_0=\frac{1}{2}(\kappa_{nn}+\kappa_{np})$, i.e.,
$\kappa_1=\alpha\kappa_0$, $\alpha$ being a fitting parameter.
Following ref. \cite{BaSc2}, we take $\alpha=-2$. For the isoscalar
strength constant we take the self consistent value (see appendix B).

This set of equations has two integrals of motion:
\begin{eqnarray}
 \frac{i\hbar}{m\Delta'}\underline {\tilde P}_-
+\left[\frac{\hbar^2\chi_0k_0'}{4m\Delta'}
-2m\bar\omega^2\psi\right]
\frac{i\hbar}{\chi_0k_4'+2\Delta'}\underline {\tilde Q}_-
-\underline L_-=\mathit{const}
\label{Int2iv}
\end{eqnarray}
and
\begin{equation}
 \underline I_-+i\hbar 2m\bar\omega^2\delta(1-\alpha)
\frac{1-\xi}{\chi_0k_4'+2\Delta'}
\underline {\tilde Q}_-
=\mathit{const}.
\label{Int1iv}
\end{equation}
By definition, the variable $\underline {\tilde Q}_-$ is purely imaginary because
$\kappa_-$ is the imaginary part of the anomalous density
$\kappa$. Therefore Eq.~(\ref{Int1iv}) implies that the relative
angular momentum $\underline I_-$ oscillates in phase with the relative
quadrupole moment $\underline {\tilde Q}_-$ of the imaginary part of
$\kappa$. Analogously, one can interpret Eq.~(\ref{Int2iv}) saying
that the variable $\underline L_-$ oscillates out of phase with the linear
combination of two variables $\underline {\tilde Q}_-$ and
$\underline {\tilde P}_-$ describing
the quadrupole deformation of the anomalous density $\kappa$ in
coordinate and momentum spaces, respectively.

Imposing the time evolution via $e^{iEt/\hbar}$ for all variables, one
transforms the equations (\ref{Ivec}) into a set of algebraic
equations, whose determinant gives the eigenfrequencies of the
system. We have
\begin{equation}
E^2\{E^4-2\D_{\omega}E^2
+8\Delta'\tilde\Delta'\hbar^2\omega^2\psi
+4(\hbar\bar{\omega})^4\delta^2(1-\alpha)(1-\xi)(1-\alpha\xi)
-\chi_0\tilde\Delta' k_0'\hbar^4/m^2
\}=0,
\label{charac1}
\end{equation}
where $\D_{\omega}=2\Delta'\tilde\Delta'+\hbar^2\bar\omega^2
[(1+\frac{\delta}{3})(2-\alpha-\xi)-\frac{2}{3}\delta\alpha\xi]$.
The solution $E=0$ corresponds to the integrals of motion
(\ref{Int2iv}) and (\ref{Int1iv}). Two nontrivial solutions of
(\ref{charac1})
\begin{equation}
E^2_{\pm}=\D_{\omega}\pm\sqrt{\D_{\omega}^2
-8\Delta'\tilde\Delta'\hbar^2\omega^2\psi
-4(\hbar\bar{\omega})^4\delta^2(1-\alpha)(1-\xi)(1-\alpha\xi)
+\chi_0\tilde\Delta' k_0'\hbar^4/m^2}
\label{EpmV}
\end{equation}
describe the energy $E_+$ of the IsoVector Giant Quadrupole Resonance
(IVGQR) and the energy $E_-$ of the scissors mode.  It is worth noting
that contrary to the case without pairing \cite{BaSc2} the energy
$E_-$ does not go to zero for deformation $\delta=0$. However, this
does not contradict the known quantum mechanical statement that the
rotation of spherical nuclei is impossible.It is easy to see from
(\ref{Int1iv}) that the relative angular momentum $\underline I_-$ is
conserved in this case, $\underline I_-=\mathit{const}$, so this mode
of a spherical nucleus has nothing in common with a vibration of angular
momentum. The calculation of transition probabilities (see below)
shows that it can be excited by an electric field and it is not
excited by a magnetic field. Our estimate of the energy of this mode
gives a value of about 2.88 MeV, which is not far from the result of
M. Matsuo et al \cite{Matsuo}, who studied the isovector quadrupole
response of $^{158}$Sn in the framework of QRPA with Skyrme forces and
found the proper resonance at $\sim$2.2 MeV.

It is known \cite{BaSc2,Zaw} that without pairing the scissors mode
lies at non-zero energy only due to Fermi Surface Deformation
(FSD). Let us investigate the role of FSD in the case with pairing.
Omitting in (\ref{Ivec}) the variable $\underline P_+$ responsible for
FSD and its dynamical equation, we obtain the characteristic equation
\begin{equation}
E^3\{E^2-
2(\hbar\omega)^2[(1+\frac{\delta}{3})(1-\alpha-\xi)
-\frac{\delta}{3}\alpha\xi]\}=0,
\end{equation}
which coincides with the analogous equation of \cite{BaSc2} derived
without pairing for $\xi=0$ ($N=Z$). The two solutions $E_{low}^2=0$
and $E_{high}^2= 2(\hbar\omega)^2[(1+\frac{\delta}{3})(1-\alpha-\xi)
-\frac{\delta}{3}\alpha\xi]$ demonstrate in an obvious way that the
role of FSD is not very important for IVGQR, whereas it is crucial for
the scissors mode and the ISGQR, whose energy in the approximation
$\xi=0$ can be obtained from the IVGQR by assuming $\alpha=1$ (see
below).

\subsection{Transition probabilities}

The transition probabilities are calculated with the help of linear
response theory. The detailed description of its use within the
framework of the WFM method can be found in \cite{BaSc2}, so we only
present the final results.

Electric quadrupole excitations are described by the operator
\begin{equation}
\label{Oelec}
\hat F=\hat F_{2\mu}^{\rm p}=\sum_{s=1}^Z\hat f_{2\mu}(s), \quad
\hat f_{2\mu}=e\,r^2Y_{2\mu}.
\end{equation}
The transition probabilities of the isovector modes are
\begin{eqnarray}
\label{E2nu}
B(E2)_{\nu}=2|<\nu|\hat F_{21}^{\rm p}|0>|^2
=\frac{e^2\hbar^2}{m}\frac{5}{4\pi}Q_{00}^{\rm p}
\frac{(1+\delta/3)(E_{\nu}^2-4\Delta'\tilde\Delta')
-2(\hbar\bar\omega\delta)^2(1-\alpha\xi)}
{E_{\nu}[E^2_{\nu}-\D_{\omega}]}.
\end{eqnarray}

Magnetic dipole excitations are described by the operator
\begin{equation}
\label{Omagn}
\hat F=\hat F_{1\mu}^{\rm p}=\sum_{s=1}^Z\hat f_{1\mu}(s), \quad
\hat f_{1\mu}=-i\nabla
(rY_{1\mu})\cdot[\br\times\nabla]\mu_N,\quad
\mu_N=\frac{e\hbar}{2mc}.
\end{equation}
Their transition probabilities are
\begin{eqnarray}
\label{scimat}
B(M1)_{\nu}=
\frac{m\bar\omega^2}{4\pi}(1-\alpha)(1-\xi)Q_{00}^{\rm p}\delta^2
\frac{E_{\nu}^2-4\Delta'\tilde\Delta'
-2(\hbar\bar\omega)^2[1+\frac{\delta}{3}(1-\alpha\xi)]}
{E_{\nu}[E^2_{\nu}-\D_{\omega}]}
\,\mu_N^2.
\end{eqnarray}

Multiplying the B(M1) factors of both states by their respective
energies and summing up, we find the following formula for the
energy-weighted sum rule
\begin{eqnarray}
\label{Msum}
E_{sc}B(M1)_{sc}+E_{iv}B(M1)_{iv}=(1-\alpha)(1-\xi)
\frac{m\bar\omega^2}{2\pi}
Q_{00}^{\rm p}\delta^2\mu_N^2.
\end{eqnarray}
The same manipulations with the B(E2) factors give
\begin{eqnarray}
\label{Esum}
E_{ISL}B(E2)_{ISL}+E_{is}B(E2)_{is}
+E_{sc}B(E2)_{sc}+E_{iv}B(E2)_{iv}
=e^2\frac{\hbar^2}{m}
\frac{5}{2\pi}
Q_{00}^{\rm p}(1+\delta/3).
\end{eqnarray}
It turns out that both expressions coincide exactly with the
respective sum rules calculated in \cite{BaSc2} without pairing. Does
this mean that there is no contribution to the sum rules which comes
from pairing?  Of course not, because the value of the mean square
radius $Q_{00}^{\rm p}$ should be calculated with the ground state
wave function which depends on pair correlations: $Q_{00}^{\rm
p}=\sum_iv_i^2\langle i|r^2|i\rangle$.

This is a good place for discussing the deformation dependence of the
energies and transition probabilities of the isovector modes. First we
recall the relevant formulae without pairing \cite{BaSc2}:
$$(E^0_{iv})^2=4\hbar^2\bar\omega^2
\left(\delta_3+\sqrt{(\delta_3^2-
 \frac{3}{4}\delta^2}\,\right),\quad
(E^0_{sc})^2=4\hbar^2\bar\omega^2
\left(\delta_3-\sqrt{\delta_3^2-
\frac{3}{4}\delta^2}\,\right),$$
\begin{eqnarray}
B(M1)^0_{\nu}=
\frac{3}{4\pi}m\bar\omega^2
Q_{00}^{\rm p}\delta^2\frac{E_{\nu}^2-2\hbar^2\bar\omega^2\delta_3}
{E_{\nu}(E^2_{\nu}-4\hbar^2\bar\omega^2\delta_3)}\,\mu_N^2,
\label{M1sc}
\end{eqnarray}
where $\delta_3=1+\delta/3$, the superscript ``0'' means the absence
of pairing and we assumed $\alpha=-2$. For the sake of simplicity we
put $\xi=0$. The scissors-mode energy is proportional to $\delta$,
which becomes evident after expanding the square root:
\begin{equation}
\label{Esc0}
E^0_{sc}=\delta\hbar\bar\omega\sqrt{\frac{3}{2\delta_3}}
\left(1+\frac{3}{16}\frac{\delta^2}{\delta_3^2}
+\frac{9}{128}\frac{\delta^4}{\delta_3^4}+...\right).
\end{equation}

At a first glance, the transition probability, as given by formula
(\ref{M1sc}), seems to have the desired (experimentally observed)
quadratic deformation dependence. However, due to the linear
$\delta$-dependence of the factor $E_{sc}$ in the denominator, the
resulting $\delta$-dependence of $B(M1)_{sc}^0$ turns out to be
linear, too. The situation is completely different when pairing is
included. In this case, the main contribution to the scissors mode
energy comes from the pairing interaction, $E_{sc}$ is not
proportional to $\delta$, and the deformation dependence of
$B(M1)_{sc}$ becomes quadratic in excellent agreement with QRPA
calculations and experimental data
\cite{Zaw,Lo2000,Magnus,Garrido,Hamam,Sushkov,Macfar,Hilt86}.

The deformation dependence of $B(M1)_{iv}$ is quadratic in $\delta$,
even without pairing, because the energy $E_{iv}$ is not proportional
to $\delta$ and depends only weakly on it. The inclusion of pairing
does not change this picture.

\subsection{Isoscalar modes}

The set of equations describing isoscalar excitations reads
\begin{eqnarray}
i\hbar {\dot Q}_+&=&i\hbar \frac{1}{m}{L}_-,
\nonumber\\
i\hbar {\dot{\tilde Q}}_-&=&
\{\chi_0[k_4^++\alpha(\nu-\pi)k_4^-]+2\Delta'\}{Q}_+,
\nonumber\\
i\hbar {\dot P}_+&=&
2\tilde\Delta'{\tilde P}_-
-i\hbar m\bar\omega^2
\{[1+\frac{\delta}{3}(1-\alpha\xi)]{L}_-
-\delta(1-\alpha\xi) {I}_-\},
\nonumber\\
i\hbar {\dot{\tilde P}}_-&=&
2\Delta'{P}_+
-\frac{\hbar^2}{4}\chi_0[k_0^++\alpha(\nu-\pi)k_0^-]{Q}_+,
\nonumber\\
i\hbar {\dot L}_-&=&i\hbar \frac{2}{m}{P}_+
+i\hbar
2m\bar\omega^2(1+\frac{2}{3}\delta)\alpha\xi {Q}_+,
\nonumber\\
i\hbar {\dot I}_-&=&0.
\label{Iscal}
\end{eqnarray}
As it is seen from the last equation, the angular momentum $
I_-=I_-^{\rm n}+I_-^{\rm p}$ is conserved, as it should be, since we
work with the rotationally invariant mean field Hamiltonian
(\ref{htau}).

Assuming for simplicity $N=Z$, we find the following characteristic
equation
\begin{equation}
E^2\{E^4-E^2[4\Delta\tilde\Delta+2\epsilon^2]
-\chi_0\tilde\Delta k_0\hbar^4/m^2\}=0,
\label{charac0}
\end{equation}
where $\epsilon^2= \hbar^2\omega^2(1+\frac{\delta}{3})$. The two solutions
of (\ref{charac0})
\begin{equation}
E^2_{\pm}=2\Delta\tilde\Delta+\epsilon^2
\pm\sqrt{(2\Delta\tilde\Delta+\epsilon^2)^2
+\chi_0\tilde\Delta k_0\hbar^4/m^2}
\label{EpmS}
\end{equation}
give the energy $E_+\equiv E_{is}$ of the IsoScalar Giant Quadrupole Resonance
(ISGQR) and the energy $E_-\equiv E_{ISL}$ of the IsoScalar Low-Lying
Excitation
(ISLLE). If one neglects the quantum correction (the term with $k_0$),
one finds for the ISGQR energy the expression
\begin{equation}
E^2_{GQR}=2\epsilon^2+4\Delta\tilde\Delta,
\label{Giant1}
\end{equation}
which is reduced to the standard value
$E_{GQR}=\sqrt2\hbar\omega(1+\frac{1}{3}\delta)$ in the case
$\Delta=0$. In this case, the energy of the low-lying mode disappears.
The transition probabilities of the isoscalar modes in the
approximation $N=Z$ are obtained from (\ref{E2nu}) by taking
$\alpha=1$.

It is important to note that the very existence of the ISLLE relies on
two factors: 1) pair correlations and 2) quantum correction. With the
parameters given in section 4.4, its energy and transition probability
for $^{164}$Dy can be estimated to be $E_{ISL}=1.0$ MeV, $B(E2)=41.4$
W.u.. These numbers are of the right order of magnitude
\cite{Solov}. Nevertheless, we do not dare to compare them with an
experiment until all terms proportional to $\hbar^2$ in
(\ref{HFBWlin}) are taken into account. The accurate calculation of
the quantum correction and the comparison of $E_{ISL}$ and
$B(E2)_{ISL}$ with experimental data will be postponed to a future
publication.

\subsection{Numerical results for the scissors mode}

We have reproduced all experimentally observed qualitative features of
the scissors mode. We understand that the harmonic oscillator model
with QQ residual interaction is too simple to give a precise
quantitative description of the experimental results. Moreover, even
within this simple model we had to make the additional approximation
(\ref{approx1}). Nevertheless let us calculate the energies and
$B(M1)$ factors to get an idea of the order of magnitude of the
discrepancy with experimental data. We will also compare our results
with those of \cite{Malov} in order to see the effect of the violation
or non-violation of the continuity equation.

Results for most of the nuclei where this mode has been observed are
presented in Table \ref{table1} and in Figures 1 -- 3.  The formulae
(\ref{EpmV}) and (\ref{scimat}) were used with the following values of
the parameters: $\alpha=-2$, $Q_{00}=A\frac{3}{5}R^2$, $R=r_0A^{1/3}$,
$r_0=1.2$ fm, $\bar\omega^2=\omega_0^2/
[(1+\frac{4}{3}\delta)^{2/3}(1-\frac{2}{3}\delta)^{1/3}]$,
$\hbar\omega_0=41/A^{1/3}$ MeV, $\hbar^2/m=41.803$ MeV fm$^2$. The gap
$\Delta$ as well as the integrals $I^{\kappa\Delta}_{pp}$ and $k_0$
were calculated with the help of the semiclassical formulae for
$\kappa(\br,p)$ and $\Delta(\br,p)$ (see Appendix A), a Gaussian being
used for the pairing interaction with $r_p=1.9$ fm and $V_0=25$
MeV. The dependence of $\Delta(\br,p_F(\br))$ and
$I^{\kappa\Delta}_{pp}(\br,p_F(\br))$ (calculated for $^{164}$Er) on
the coordinate $r$ is demonstrated on fig. 1. We checked that the
dependence of $\Delta(\br,p_F(\br))$,
$I^{\kappa\Delta}_{pp}(\br,p_F(\br))$ and $p_F(\br)$ on the direction
of the vector $\br$ is negligible. The best agreement of the theory
with the experiment is obtained at the point $\bar r$ where the
integral $I^{\kappa\Delta}_{pp}(\br,p_F(\br))$ has its maximum.  The
values of $k_0^+,$ vary smoothly from $k_0^+=62.5$ for $A=134$ to
$k_0^+=72.2$ for $A=196$. The analysis of Table 1 shows that the
overall agreement of the theoretical results with the experimental
data is reasonable. It is, of course, not perfect, but the influence
of pairing, especially on the $B(M1)$ values, is impressive.  Without
pairing, the calculated energies (column $\Delta=0$) are 1 -- 2 MeV
(1.5 -- 2 times) smaller than $E_{exp}$, and the $B(M1)$ factors
(column $\Delta=0$) are 3--7 times larger than the experimental
values.  The inclusion of pairing changes the results (columns $new$)
drastically: the discrepancy between calculated and experimental
energies is reduced to 5 -- 20$\%$ and the calculated transitions
probabilities are reduced by a factor of 1.5 -- 3.  The influence of
$k_0$ is negligibly small, being of the order of $\sim 2\%$.
\begin{table}
\caption{\small Scissors mode energies $E_{sc}$ (in MeV) and
transition probabilities $B(M1)_{sc}$ (in units of $\mu_N^2$); $\mathit{exp}$:
experimental values, $\mathit{old}$: old theory \cite{Malov} with violated
continuity equation, $\mathit{new}$: new theory with fulfilled continuity
equation, $\Delta=0$: theory without pairing. The experimental values
of $E_{sc}$, $\delta$ and $B(M1)$ are from Ref. \cite{Pietr} and
references therein.}
\label{table1}
\vspace{-0.2cm}
\begin{center}
\begin{tabular}{|c|c||c|c|c|c||c|c|c|c|}
\hline
   & &
\multicolumn{4}{|c||}{ $E_{sc}$}    &
\multicolumn{4}{|c|}{ $B(M1)_{sc}$}    \\
\cline{3-10}
 Nuclei & $\delta$ & $\mathit{exp}$ & $\mathit{old}$ & $\mathit{new}$
 & $\Delta=0$ & $\mathit{exp}$ & $\mathit{old}$ & $\mathit{new}$ & $\Delta=0$ \\
\hline
 $^{134}$Ba & 0.14 & 2.99 & 3.94 & 3.09 & 1.28 & 0.56 & 1.16 & 1.67 & 3.90 \\[-2mm]
 $^{144}$Nd & 0.11 & 3.21 & 3.86 & 3.03 & 1.04 & 0.17 & 0.86 & 1.25 & 3.54 \\[-2mm]
 $^{146}$Nd & 0.13 & 3.47 & 3.91 & 3.09 & 1.18 & 0.72 & 1.13 & 1.62 & 4.14 \\[-2mm]
 $^{148}$Nd & 0.17 & 3.37 & 4.02 & 3.18 & 1.48 & 0.78 & 1.79 & 2.58 & 5.39 \\[-2mm]
 $^{150}$Nd & 0.22 & 3.04 & 4.25 & 3.44 & 1.92 & 1.61 & 2.94 & 4.17 & 7.26 \\[-2mm]
 $^{148}$Sm & 0.12 & 3.07 & 3.88 & 3.00 & 1.11 & 0.43 & 1.02 & 1.50 & 3.96 \\[-2mm]
 $^{150}$Sm & 0.16 & 3.13 & 4.00 & 3.13 & 1.42 & 0.92 & 1.68 & 2.45 & 5.26 \\[-2mm]
 $^{152}$Sm & 0.24 & 2.99 & 4.30 & 3.46 & 2.02 & 2.26 & 3.27 & 4.68 & 7.81 \\[-2mm]
 $^{154}$Sm & 0.26 & 3.20 & 4.39 & 3.57 & 2.17 & 2.18 & 3.79 & 5.42 & 8.65 \\[-2mm]
 $^{156}$Gd & 0.26 & 3.06 & 4.39 & 3.60 & 2.16 & 2.73 & 3.82 & 5.42 & 8.76 \\[-2mm]
 $^{158}$Gd & 0.26 & 3.14 & 4.41 & 3.60 & 2.19 & 3.39 & 4.01 & 5.72 & 9.12 \\[-2mm]
 $^{160}$Gd & 0.27 & 3.18 & 4.42 & 3.61 & 2.21 & 2.97 & 4.14 & 5.90 & 9.38 \\[-2mm]
 $^{160}$Dy & 0.26 & 2.87 & 4.37 & 3.59 & 2.13 & 2.42 & 3.89 & 5.53 & 9.03 \\[-2mm]
 $^{162}$Dy & 0.26 & 2.96 & 4.38 & 3.61 & 2.14 & 2.49 & 3.99 & 5.66 & 9.25 \\[-2mm]
 $^{164}$Dy & 0.26 & 3.14 & 4.40 & 3.60 & 2.17 & 3.18 & 4.17 & 5.95 & 9.59 \\[-2mm]
 $^{164}$Er & 0.25 & 2.90 & 4.35 & 3.57 & 2.10 & 1.45 & 3.94 & 5.62 & 9.26 \\[-2mm]
 $^{166}$Er & 0.26 & 2.96 & 4.37 & 3.53 & 2.13 & 2.67 & 4.12 & 5.96 & 9.59 \\[-2mm]
 $^{168}$Er & 0.26 & 3.21 & 4.36 & 3.53 & 2.10 & 2.82 & 4.11 & 5.95 & 9.67 \\[-2mm]
 $^{170}$Er & 0.26 & 3.22 & 4.35 & 3.57 & 2.09 & 2.63 & 4.14 & 5.91 & 9.79 \\[-2mm]
 $^{172}$Yb & 0.25 & 3.03 & 4.33 & 3.55 & 2.05 & 1.94 & 4.08 & 5.84 & 9.79 \\[-2mm]
 $^{174}$Yb & 0.25 & 3.15 & 4.31 & 3.47 & 2.02 & 2.70 & 4.05 & 5.89 & 9.82 \\[-2mm]
 $^{176}$Yb & 0.24 & 2.96 & 4.26 & 3.45 & 1.94 & 2.66 & 3.83 & 5.54 & 9.58 \\[-2mm]
 $^{178}$Hf & 0.22 & 3.11 & 4.19 & 3.43 & 1.79 & 2.04 & 3.40 & 4.86 & 9.00 \\[-2mm]
 $^{180}$Hf & 0.22 & 2.95 & 4.17 & 3.36 & 1.76 & 1.61 & 3.34 & 4.85 & 8.97 \\[-2mm]
 $^{182}$W  & 0.20 & 3.10 & 4.10 & 3.30 & 1.63 & 1.65 & 2.96 & 4.31 & 8.43 \\[-2mm]
 $^{184}$W  & 0.19 & 3.31 & 4.07 & 3.28 & 1.55 & 1.12 & 2.74 & 3.97 & 8.14 \\[-2mm]
 $^{186}$W  & 0.18 & 3.20 & 4.04 & 3.26 & 1.49 & 0.82 & 2.60 & 3.76 & 7.95 \\[-2mm]
 $^{190}$Os & 0.15 & 2.90 & 3.93 & 3.12 & 1.21 & 0.98 & 1.82 & 2.67 & 6.64 \\[-2mm]
 $^{192}$Os & 0.14 & 3.01 & 3.90 & 3.12 & 1.15 & 1.04 & 1.66 & 2.42 & 6.37 \\[-2mm]
 $^{196}$Pt & 0.11 & 2.68 & 3.83 & 3.01 & 0.94 & 0.70 & 1.16 & 1.72 & 5.35 \\
\hline
\end{tabular}
\end{center}
\end{table}

The influence of the correct treatment of the continuity equation can
be estimated by comparing the columns $\mathit{new}$ (results with the fulfilled
continuity equation) and $\mathit{old}$ (results from ref. \cite{Malov} with the
violated continuity equation). It turns out that the correct treatment
of the continuity equation has a substantial effect on the results: on
the one hand, it leads to a decrease of the scissors-mode energies by
0.8 -- 0.9 MeV (in comparison with the old results), improving the
agreement with the experimental data, but on the other hand, it
results in an increase of the transition probabilities by a factor 1.4
-- 1.5, deteriorating the agreement with experiment.

What can be done to improve these results? An obvious idea is to get
rid off the approximation (\ref{approx1}), i.e., to solve the full set
of eight equations (\ref{HFB8}), calculating all integrals within a
microscopic approach. The next possible step is to perform a
self-consistent calculation with a more or less realistic interaction.

\begin{figure}
\begin{center}
\epsfig{file=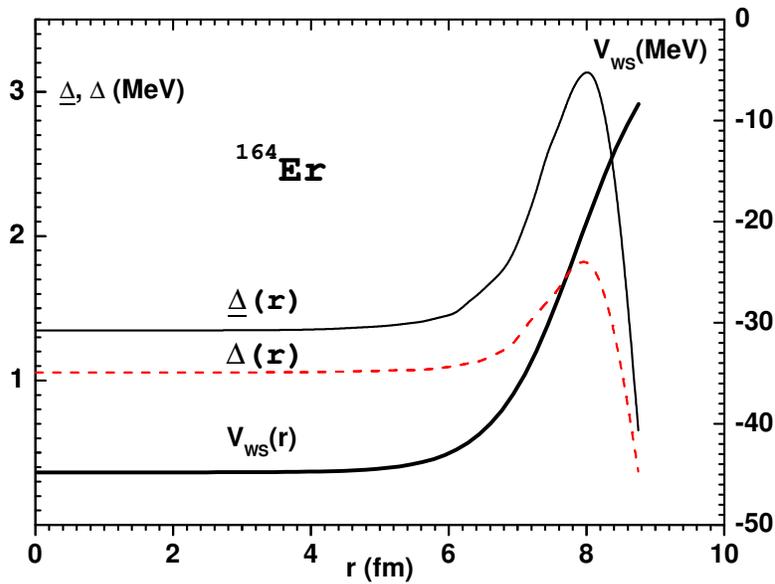,width=12cm}
\end{center}
\caption{\small The pair field $\Delta(r,p_F(r))$, the value of
$\underline{\Delta}=2\tilde\Delta=|V_0|I^{\kappa\Delta}_{pp}(r,p_F(r))$
and the
Woods-Saxon potential $V_{WS}(r)$ as the functions of radius $r$.}
\label{figure1}
\end{figure}

\begin{figure}
\begin{center}
\epsfig{file=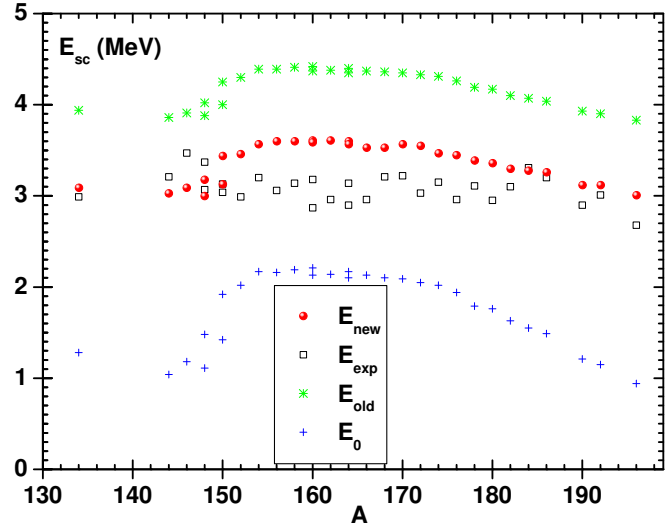,width=10cm}
\end{center}
\caption{\small Scissors mode energies as a function of the mass
number $A$ for the nuclei listed in Table 1. $E_{new}$: new theory,
$E_{old}$: old theory \cite{Malov}, $E_0$: theory without pairing.}
\label{figure2}
\end{figure}

\begin{figure}
\begin{center}
\epsfig{file=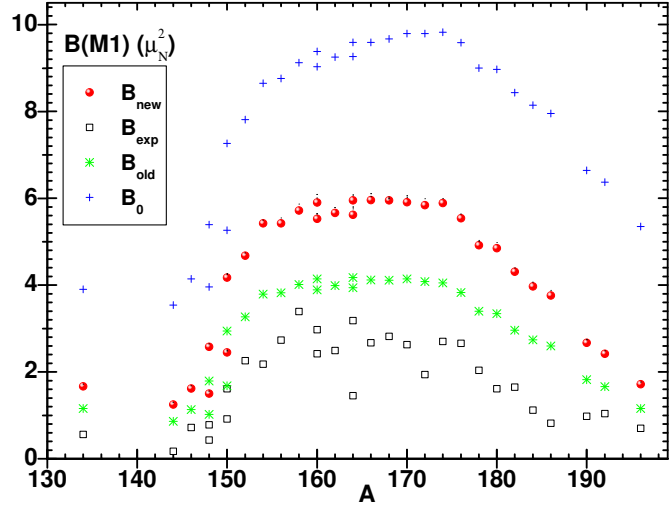,width=10cm}
\end{center}
\caption{\small Scissors mode transition probabilities $B(M1)$ as a
function of the mass number $A$ for the nuclei listed in Table 1.
$B_{new}$: new theory, $B_{old}$: old theory \cite{Malov}, $B_0$:
theory without pairing.}
\label{figure3}
\end{figure}

Another point which should be clarified is the role of the spin-orbit
interaction. It is known \cite{Zaw} that the experimentally observed
low-lying magnetic dipole strength consists of two separated parts:
orbital excitations in the energy interval $\sim$2 -- 4 MeV and the
spin-flip resonance ranging from 5 to 10 MeV excitation energy. So,
for the full description of the scissors mode it would be important to
consider also the spin degrees of freedom.

\section{Conclusion}

In conclusion, we presented a generalization of the method of Wigner
function moments which allowed us to include pair correlations without
violating the continuity equation. The method was exemplified by the
calculation of isovector and isoscalar giant resonances and low-lying
excitations in the harmonic oscillator model with
quadrupole-quadrupole residual interaction.

The analytical formulae, derived in a slightly simplified model
(approximation (\ref{approx1})), reproduce very well the experimentally
observed deformation dependence of the energy and $B(M1)$ factor of
the scissors mode. We performed calculations for most of the nuclei
where this mode has been observed, and our results are in reasonable
quantitative agreement with the experimental data, the pair correlations
being extremely important.

The joint action of pairing and quantum corrections lead to the
appearance of a low-lying isoscalar excitation. The low-lying
isoscalar, as well as isovector, modes are very sensitive to the
parameters of the interaction and to all details of the ground-state
input. Their study on the basis of the full (8 variables) set of
dynamical equations (\ref{HFB8}), with normal and anomalous densities
calculated within a microscopic approach, are in progress.

\section*{Acknowledgements}

The discussions with A. Severyukhin are gratefully acknowledged.

\section*{Appendix A}

Let us consider integrals containing $\delta \kappa$ and
$\delta\Delta$.

The relevant term of the first equation in (\ref{HFBWlin}) with
$W=p_xp_z$ reads
\begin{eqnarray}
S=2\int\! d(\bp,\br)\,
p_xp_z(\Delta^{0}\delta\kappa_--\kappa^{0}\delta \Delta_-).
\label{Int1}
\end{eqnarray}
Substituting $\Delta$ and $\delta\Delta$ according to the expressions
(\ref{DK}) and (\ref{varDK}), one gets
\begin{equation}
S=-\int\! d^3r
\int\! \frac{2d^3p}{(2\pi\hbar)^3}
\int\! \frac{2d^3p'}{(2\pi\hbar)^3}
p_xp_zv(|\bp-\bp'|)[\kappa^0(\br,\bp')\delta\kappa_-(\br,\bp,t)
-\kappa^0(\br,\bp)\delta \kappa_-(\br,\bp',t)].
\end{equation}
Changing in the second part of the integral the variables
$p\leftrightarrow p'$, one finds
\begin{eqnarray}
S=-\int\! d^3r
\int\! \frac{2d^3p}{(2\pi\hbar)^3}
\int\! \frac{2d^3p'}{(2\pi\hbar)^3}
v(|\bp-\bp'|)\kappa^0(\br,\bp')[p_xp_z-p'_xp'_z]
\delta \kappa_-(\br,\bp,t).
\label{I1}
\end{eqnarray}
Let us analyze in detail the integral over $\bp'$
\begin{eqnarray}
\int_{p'}
&=&\int\!\frac{2d^3p'}{(2\pi\hbar)^3}
v(|\bp-\bp'|)\kappa^0(\br,\bp')[p_xp_z-p'_xp'_z]
\nonumber\\
&=&p_xp_z\int\!\frac{2d^3p'}{(2\pi\hbar)^3}
v(|\bp-\bp'|)\kappa^0(\br,\bp')
-\int\!\frac{2d^3p'}{(2\pi\hbar)^3}
v(|\bp-\bp'|)\kappa^0(\br,\bp')p'_xp'_z.
\label{Ip'1}
\end{eqnarray}
Now it is necessary to clarify the $\bp$ dependence of $\kappa^0$.
There are good reasons to assume that it depends only on $|\bp|=p$,
i.e., it does not depend on the angles in the momentum space. In fact,
according to the semiclassical formula \cite{Ring}
\begin{equation}
\kappa(\br,\bp)=\frac{1}{2}
\frac{\Delta(\br,\bp)}{\sqrt{h^2(\br,\bp)+\Delta^2(\br,\bp)}}
\label{kappa}
\end{equation}
the anomalous density can depend on the direction of $\bp$ only via
$\Delta(\br,\bp)$. However, as can be seen from the semiclassical gap
equation \cite{Ring}
\begin{equation}
\Delta(\br,\bp)=-\frac{1}{2}\int\!\frac{d^3p'}{(2\pi\hbar)^3}
v(|\bp-\bp'|)
\frac{\Delta(\br,\bp')}{\sqrt{h^2(\br,\bp')+\Delta^2(\br,\bp')}}\,,
\end{equation}
there are no reasons to introduce the angular dependence (in momentum
space) of $\Delta$ as long as the Fermi surface is spherical. So, it
is quite natural to take $\Delta(\br,\bp)=\Delta(\br,p)$ and, as a
consequence, $\kappa(\br,\bp)=\kappa(\br,p)$. In this case the
integral $\int_{p'}$ can be integrated over angles analytically. To
this end we expand the pairing force in a standard way
\cite{Varshal,Ring}:
\begin{equation}
v(|\bp-\bp'|)=\sum_{l=0}^{\infty}v_l(p,p')\sum_m
Y^*_{lm}(\theta,\phi)Y_{lm}(\theta',\phi').
\label{ExpanV}
\end{equation}
Having in mind that
$p_xp_z=\sqrt{\frac{15}{2\pi}} p^2(Y_{2-1}-Y_{21})$ we find
\begin{equation}
\int_{p'}
=2p_xp_z\int\!\frac{p'^2dp'}{(2\pi\hbar)^3}
\kappa^0(\br,p')v_0(p,p')
-2\sqrt{\frac{15}{2\pi}}[Y_{2-1}(\theta,\phi)-Y_{21}(\theta,\phi)]
\int\!\frac{p'^4dp'}{(2\pi\hbar)^3}\kappa^0(\br,p')v_2(p,p').
\end{equation}
If we use a Gaussian as pairing interaction \cite{Ring}:
\begin{equation}
v(|\bp-\bp'|)=\beta e^{-\alpha |\bp-\bp'|^2}\,,
\label{Gaus}
\end{equation}
with $\beta=-|V_0|(r_p\sqrt{\pi})^3$ and $\alpha=r_p^2/4\hbar^2$,
the expansion coefficients of the force read \cite{Varshal}:
\begin{equation}
v_l(p,p')=4\pi\beta e^{-\alpha(p^2+p'^2)}i^lj_l(-ix),
\label{ExpanG}
\end{equation}
where $j_l(-ix)$ is the spherical Bessel function and $x=2\alpha pp'$.
We need the first three coefficients
\begin{equation}
v_0(p,p')=4\pi\beta e^{-\alpha(p^2+p'^2)}\phi_0(x),
\quad \phi_0(x)=\frac{1}{x}sh(x)
=(1+\frac{x^2}{6}+\dots),
\end{equation}
\begin{equation}
v_1(p,p')=4\pi\beta e^{-\alpha(p^2+p'^2)}x\phi_1(x),
\quad \phi_1(x)=\frac{1}{x^2}[ch(x)-\frac{1}{x}sh(x)]
=\frac{1}{3}(1+\frac{x^2}{10}+\dots),
\end{equation}
\begin{equation}
v_2(p,p')=4\pi\beta e^{-\alpha(p^2+p'^2)}x^2\phi_2(x),
\quad \phi_2(x)=\frac{1}{x^3}[(1+\frac{3}{x^2})sh(x)-\frac{3}{x}ch(x)]
=\frac{1}{15}(1+\frac{x^2}{14}+\dots).
\end{equation}
With the help of these expressions one finds
\begin{eqnarray}
\int_{p'}=-p_xp_z|V_0|I^{\kappa\Delta}_{pp}(\br,p),
\label{Ip'}
\end{eqnarray}
where
\begin{eqnarray}
I^{\kappa\Delta}_{pp}(\br,p)=
\frac{r_p^3}{\sqrt{\pi}\hbar^3}e^{-\alpha p^2}
\int\!\kappa^0(\br,p')\left[\phi_0(2\alpha pp')
-4\alpha^2p'^4\phi_2(2\alpha pp')\right]
e^{-\alpha p'^2}p'^2dp' .
\label{Ikap}
\end{eqnarray}
Substituting  (\ref{Ip'}) into (\ref{I1}) one gets finally
\begin{eqnarray}
S&=&
|V_0|\int\! d(\bp,\br)
I^{\kappa\Delta}_{pp}(\br,p)
p_xp_z\delta \kappa_-(\br,\bp,t)
\nonumber\\
&\simeq&|V_0|I^{\kappa\Delta}_{pp}(\bar{\br},p_F)
\int\! d(\bp,\br)p_xp_z\delta \kappa_-(\br,\bp,t)=
|V_0|I^{\kappa\Delta}_{pp}(\bar{\br},p_F)\tilde P_-(t).
\label{I11}
\end{eqnarray}

The third and fourth equations of (\ref{HFBWlin}) contain the terms
\begin{equation}
T_{\pm}^W=\int\! d(\bp,\br)\,
W(1-f_+^0)\delta \Delta_{\pm}.
\label{Ifd}
\end{equation}
As above, we assume that $\Delta$ does not depend on the
direction of $\bp$. On the basis of the semiclassical formula
\begin{equation}
1-f_+^0(\br,\bp)=\frac{h^0(\br,\bp)}
{\sqrt{h^2(\br,\bp)+\Delta^2(\br,\bp)}}
\end{equation}
we can assume $f_+^0(\br,\bp)=f_+^0(\br,p)$, in agreement with the
assumption of a spherical Fermi surface.  Calculating (\ref{Ifd}) with
$W=xz$, $p_xp_z$, $\hat L$ and $\hat I_y$ we obtain
\begin{equation}
T_{\pm}^{xz}=
|V_0|I^{f\Delta}_{xz}(\bar\br,p_F)\tilde Q_{\pm}(t),
\quad T_{\pm}^{p_xp_z}=
|V_0|I^{f\Delta}_{pp}(\bar\br,p_F)\tilde P_{\pm}(t),
\end{equation}
\begin{equation}
T_{\pm}^{L}=
|V_0|I^{f\Delta}_{xp}(\bar\br,p_F)\tilde L_{\pm}(t),
\quad T_{\pm}^I=
|V_0|I^{f\Delta}_{xp}(\bar\br,p_F)\tilde I_{\pm}(t),
\end{equation}
where
\begin{equation}
I^{f\Delta}_{xz}(\br,p)=
\frac{r_p^3}{2\sqrt{\pi}\hbar^3}e^{-\alpha p^2}
\int\![1-f_+^0(\br,p')]\phi_0(2\alpha pp')e^{-\alpha p'^2}p'^2dp',
\end{equation}
\begin{equation}
I^{f\Delta}_{xp}(\br,p)=
\frac{r_p^5}{4\sqrt{\pi}\hbar^5}e^{-\alpha p^2}
\int\![1-f_+^0(\br,p')]\phi_1(2\alpha pp')e^{-\alpha p'^2}p'^4dp' .
\end{equation}
\begin{equation}
I^{f\Delta}_{pp}(\br,p)=
\frac{r_p^7}{8\sqrt{\pi}\hbar^7}e^{-\alpha p^2}
\int\![1-f_+^0(\br,p')]\phi_2(2\alpha pp')e^{-\alpha p'^2}p'^6dp',
\end{equation}
The second and third equations of (\ref{HFBWlin}) contain the terms
\begin{eqnarray}
K^W=2\int\! d(\bp,\br)\,
(\{W,\Delta^{0}\}\delta \kappa_+
-\{W,\kappa^{0}\}\delta \Delta_+)
\label{Int2}
\end{eqnarray}
and
\begin{equation}
G^W=
\int\! d(\bp,\br)\,\{W,\Delta_+^0\}\delta f_-,
\label{Ifdel}
\end{equation}
which require the knowledge of derivatives of $\Delta^0(\br,\bp)$.  They
are found with the help of formula (\ref{DK}). The derivative with
respect to $\bp$ reads:
\begin{equation}\frac{\partial\Delta^0}{\partial p_i}=-\frac{|V_0|}{2\hbar^2}
I_{\partial p}(\br,p)p_i,
\end{equation}
where
\begin{eqnarray}
I_{\partial p}(\br,p)=
\frac{r_p^5}{2\sqrt{\pi}\hbar^3}e^{-\alpha p^2}
\int\!\kappa^0(\br,p')\left[\phi_0(2\alpha pp')
-2\alpha p'^2\phi_1(2\alpha pp')\right]
e^{-\alpha p'^2}p'^2dp' .
\label{Idp}
\end{eqnarray}
To calculate the derivative with respect to $\br$, we approximate
$\kappa^0(\br,p)$ by formula (\ref{kappa}). As a result, we obtain an
integral equation for $\partial\Delta^0(\br,p)/\partial x_i$, with a
kernel which is strongly peaked at $p'=p_F$, which allows us to
simplify the equation by replacing $\partial\Delta^0(\br,p')/\partial
x_i$ under the integral by $\partial\Delta^0(\br,p_F)/\partial
x_i$. Finally we have
\begin{equation}
\frac{\partial\Delta^0(\br,p)}{\partial x_i}=\beta(\br,p,p_F)
m\omega_i^2x_i,
\end{equation}
where
\begin{equation}
\beta(\br,p,p_F)=\int\!\frac{p'^2dp'}{(2\pi\hbar)^3}
\left[v_0(p,p')-v_0(p_F,p')\frac{\gamma(\br,p)}{1+\gamma(\br,p_F)}
\right]\frac{h(\br,p')\Delta(\br,p')}{2[h^2(\br,p')
+\Delta^2(\br,p')]^{3/2}}
\end{equation}
and
\begin{equation}
\gamma(\br,p)=\int\!\frac{p'^2dp'}{(2\pi\hbar)^3}v_0(p,p')
\frac{h^2(\br,p')}{2[h^2(\br,p')
+\Delta^2(\br,p')]^{3/2}}.
\end{equation}
Calculating now (\ref{Ifdel}) with $W=xz$, we find
\begin{equation}
G^{xz}=
2\int\! d(\bp,\br)\,(z\frac{\partial\Delta^0}{\partial p_x}
+x\frac{\partial\Delta^0}{\partial p_z})\delta f_-(\br,\bp,t)=
-\frac{|V_0|}{\hbar^2}
I_{\partial p}(\bar{\br},p_F)L_-(t).
\label{Ifdel2}
\end{equation}
The calculation of (\ref{Ifdel}) with $W=p_xp_z$ gives
\begin{equation}
G^{pp}=
-2\int\! d(\bp,\br)\,(p_z\frac{\partial\Delta^0}{\partial x}
+p_x\frac{\partial\Delta^0}{\partial z})\delta f_-(\br,\bp,t)=
|V_0|I_{\partial r}(\bar{\br},p_F)[\omega_+^2L_-(t)-\omega_-^2I_-(t)],
\label{Ifdel3}
\end{equation}
where
\begin{eqnarray}
I_{\partial r}(\br,p)=
\frac{mr_p^3}{4\sqrt{\pi}\hbar^3}e^{-\alpha p^2}
\int\!\phi_0(2\alpha pp')
\frac{h(\br,p')\Delta(\br,p')-h^2(\br,p')\beta(\br,p',p_f)}
{[h^2(\br,p')+\Delta^2(\br,p')]^{3/2}}
e^{-\alpha p'^2}p'^2dp' .
\label{Idr}
\end{eqnarray}

Calculating (\ref{Int2}) with the weight $xp_z$, one gets
\begin{eqnarray}
K^{xp_z}=
-\frac{|V_0|}{\hbar^2}I^{\{\kappa\Delta\}}_{xp}(\bar\br,p_F)
\tilde P_+(t),
\label{Ixpz}
\end{eqnarray}
with
\begin{eqnarray}
I^{\{\kappa\Delta\}}_{xp}(\br,p)=
\frac{r_p^5}{2\sqrt{\pi}\hbar^3}e^{-\alpha p^2}
\int\!\kappa^0(\br,p')[\phi_0(x)-4\alpha p'^2\phi_1(x)+
4\alpha^2p'^4\phi_2(x)]e^{-\alpha p'^2}p'^2dp',
\label{grp1}
\end{eqnarray}
where $x=2\alpha pp'$. It is evident that $K^{xp_z}=
K^{zp_x}$. As a result
\begin{eqnarray}
K^L=
-2\frac{|V_0|}{\hbar^2}I^{\{\kappa\Delta\}}_{xp}(\bar\br,p_F)
\tilde P_+(t),\quad
K^I=0.
\label{IL}
\end{eqnarray}

\section*{Appendix B}

The necessary Poisson brackets are
\begin{equation}
\{xz,h_0^{\tau}\}=\frac{1}{m}\hat L,\quad
\{p_xp_z,h_0^{\tau}\}=-m(\omega_z^{\tau2}zp_x+\omega_x^{\tau2}xp_z)
=-\frac{m}{2}
[(\omega_z^{\tau2}+\omega_x^{\tau2})\hat L
-(\omega_x^{\tau2}-\omega_z^{\tau2})\hat I_y],
\end{equation}
\begin{equation}
\{\hat L,h_0^{\tau}\}=\frac{2}{m}p_xp_z-m(\omega_x^{\tau2}
+\omega_z^{\tau2})xz,
\quad \{\hat I_y,h_0^{\tau}\}=m(\omega_z^{\tau2}-\omega_x^{\tau2})xz,
\quad\{xz,h_1^{\tau}\}=0,\quad
\end{equation}
\begin{equation}
\{p_xp_z,h_1^{\tau}\}=
2\delta B_{20}^{\tau}(t)(xp_z-2zp_x)
-6[\beta_{11}^{\tau}(t)-\beta_{22}^{\tau}(t)]xp_z$$
$$-12[\beta_{13}^{\tau}(t)(zp_z+xp_x)+\beta_{12}^{\tau}(t)yp_z
+\beta_{23}^{\tau}(t)yp_x],
\end{equation}
\begin{equation}\{\hat L,h_1^{\tau}\}=
-2\delta B_{20}^{\tau}(t)xz
-12[\beta_{13}^{\tau}(t)(z^2+x^2)+\beta_{12}^{\tau}(t)yz
+\beta_{23}^{\tau}(t)yx]
-6[\beta_{11}^{\tau}(t)-\beta_{22}^{\tau}(t)]xz,
\end{equation}
\begin{equation}
\{\hat I_y,h_1^{\tau}\}=
6\delta B_{20}^{\tau}(t)xz
+12[\beta_{23}^{\tau}(t)yx
-\beta_{12}^{\tau}(t)yz
-\beta_{13}^{\tau}(t)(z^2-x^2)]
-6[\beta_{11}^{\tau}(t)-\beta_{22}^{\tau}(t)]xz,
\end{equation}
\begin{equation}
\{\{W,h_0^{\tau}\}\}=0,\;
\{\{xz,h_1^{\tau}\}\}=0,\;
\{\{p_xp_z,h_1^{\tau}\}\}=24\beta_{13}^{\tau}(t),\;
\{\{\hat L,h_1^{\tau}\}\}=0, \; \{\{\hat I_y,h_1^{\tau}\}\}=0.
\end{equation}
Here
\begin{equation}
\beta_{ij}^{\rm n}=\kappa J_{ij}^{\rm n}+\bar\kappa J_{ij}^{\rm p},
\quad
\beta_{ij}^{\rm p}=\kappa J_{ij}^{\rm p}+\bar\kappa J_{ij}^{\rm n},
\end{equation}
\begin{equation}
J_{ij}^{\tau}(t)=\int\!d(\bp,\br)x_ix_j\delta f^{\tau}(\br,\bp,t),\quad
J_{13}^{\tau}(t)=\frac{1}{2}Q_+^{\tau}(t),
\quad
\delta B_{20}^{\tau}(t)=2\beta_{33}^{\tau}(t)-\beta_{11}^{\tau}(t)
-\beta_{22}^{\tau}(t).
\end{equation}
Using the definition \cite{BM} $Q_{20}=\frac{4}{3}Q_{00}\delta$ and
the formula $q_{20}=2z^2-x^2-y^2$, one finds from (\ref{h01})
\begin{equation}
\omega_x^{\tau2}=\omega^2(1-\frac{2}{m\omega^2}B_{20}^{\tau}),\quad
\omega_x^{\tau2}=\omega^2(1-\frac{2}{m\omega^2}B_{20}^{\tau}).
\end{equation}
These oscillator frequencies depend on the strength constants
$\kappa_{\rm nn}$ and $\kappa_{\rm np}$. In the isoscalar case the constant
$\kappa_{0}=\frac{1}{2}(\kappa_{\rm nn}+\kappa_{\rm np})$ is fixed by the
self-consistency condition \cite{BM}
$\omega_x^2\langle x^2\rangle=
\omega_y^2\langle y^2\rangle=
\omega_z^2\langle z^2\rangle$, which allows one to find for $\kappa_0$
and the oscillator frequencies of the isoscalar field the following
expressions \cite{BaSc3}:
\begin{equation}
\kappa_0=-\frac{m\bar\omega^2}{4Q_{00}}, \quad
\omega_x^2=\frac{1}{2}(\omega_x^{\rm n2}+\omega_x^{\rm p2})
=\bar\omega^2(1+\frac{4}{3}\delta), \quad
\omega_z^2=\frac{1}{2}(\omega_z^{\rm n2}+\omega_z^{\rm p2})
=\bar\omega^2(1-\frac{2}{3}\delta),
\end{equation}
where $\bar\omega^2=\omega^2/(1+\frac{2}{3}\delta)
=\omega_0^2/[(1+\frac{4}{3}\delta)^{\frac{2}{3}}
(1-\frac{2}{3}\delta)^{\frac{1}{3}}],\,$ and
$\hbar\omega_0=41/A^{1/3}$MeV.

\end{document}